\documentclass[prd,twocolumn,showpacs,floatfix,amsmath,nofootinbib,amssymb,floatfix]{revtex4}
\usepackage{graphicx,color,dcolumn,booktabs,bm}
\usepackage{longtable,lscape}
\usepackage{txfonts}
\usepackage{overpic}
\usepackage{amssymb}
\usepackage{indentfirst}
\usepackage{feynmf}   
\usepackage{slashed}  
\usepackage{cases}
\usepackage{color}
\usepackage{multirow}
\usepackage{epstopdf}
\usepackage{enumerate}
\usepackage{graphicx,color,dcolumn,booktabs,bm}
\usepackage[colorlinks, citecolor=blue,anchorcolor=red,menucolor=red, linkcolor=red,filecolor=red,urlcolor=blue,frenchlinks=red]{hyperref}

\graphicspath{{Figures/}} %

\begin{document}

\title{Combined study of $2S$ and $1D$ open-charm mesons with natural spin-parity}
\author{Bing Chen$^{1,3}$}\email{chenbing@shu.edu.cn}
\author{Xiang Liu$^{2,3}$\footnote{Corresponding author.}}\email{xiangliu@lzu.edu.cn}
\author{Ailin Zhang$^{4}$}\email{zhangal@staff.shu.edu.cn}
\affiliation{$^1$Department of Physics, Anyang Normal University,
Anyang 455000, China\\$^2$School of Physical Science and Technology,
Lanzhou University,
Lanzhou 730000, China\\
$^3$Research Center for Hadron and CSR Physics, Lanzhou University
$\&$ Institute of Modern Physics of CAS,
Lanzhou 730000, China\\
$^4$Department of Physics, Shanghai University, Shanghai 200444,
China}


\begin{abstract}
In this paper, we perform a combined study of $2S$ and $1D$
open-charm mesons with natural spin-parity. Our results indicate
that $D^*_1(2600)/D^*_{s1}(2700)$ and  $D^*_1(2760)/D^*_{s1}(2860)$
are predominantly the $2^3S_1$ and $1^3D_1$ charmed/charmed-strange
mesons, respectively, while $D_3^*(2760)/D_{s3}^*(2860)$ can be
regarded as the $1^3D_3$ charmed/charmed-strange mesons. In
addition, some typical ratios of partial widths of the discussed
natural states are predicted, by which future experiments can test
these assignments, especially for the $2S$-$1D$ mixing scheme
existing in $D^*_1(2600)/D^*_1(2760)$ and
$D^*_{s1}(2700)/D^*_{s1}(2860)$.

\end{abstract}
 \maketitle

\section{Introduction}

Among these hadronic states, the heavy-light mesons, which contain
valence heavy quark $Q$ and light antiquark $\bar{q}$, are the
special system due to the existence of chiral symmetry and heavy
quark dynamics. The investigation of heavy-light mesons can improve
our understanding of nonperturbative quantum chromodynamics (QCD).

\begin{table*}[htbp]
\caption{The experimental information of these observed charmed and
charmed-strange mesons in the mass range 2.6 $-$ 2.9
GeV~\cite{delAmoSanchez:2010vq, Aaij:2012pc, Aaij:2013sza,
Brodzicka:2007aa, Lees:2014abp, Aubert:2006mh, Aubert:2009ah,
Aaij:2015sqa, Aaij:2014baa, Aaij:2015vea, Aaij:2014xza}. Two states
marked by $\dag$ are also named as the corresponding $D^*_J(2650)$
and $D_J(2740)$ in Ref.~\cite{Aaij:2013sza}.} \label{table1}
\renewcommand\arraystretch{1.5}
\begin{tabular*}{170mm}{l@{\extracolsep{\fill}}ccccccc}
\toprule[1pt]\toprule[1pt]
State                      &    Production                        & Observed decays                  &    Mass  (MeV)                   &    Width (MeV)          &     $J^P$      &      Year  &    Collaboration           \\
\midrule[1pt]
 $D^*_{1}(2600)$$^\dag$    & $e^+e^-$                             & $D^+\pi^-$, $D^{*+}\pi^-$       &  2608.7$\pm$2.4$\pm$2.5       &    93$\pm$6$\pm$13          &                & 2010       & \emph{BABAR}~\cite{delAmoSanchez:2010vq}   \\
                           &  $pp$                                & $D^{*+}\pi^-$                   &  2649.2$\pm$3.5$\pm$3.5       &    140.2$\pm$17.1$\pm$18.6  &                & 2013       & LHCb~\cite{Aaij:2013sza}  \\
 $D(2750)$$^\dag$        & $e^+e^-$                             & $D^{*+}\pi^-$                   &  2752.4$\pm$1.7$\pm$2.7       &    71$\pm$6$\pm$11          &                & 2010       & \emph{BABAR}~\cite{delAmoSanchez:2010vq}   \\
                           &  $pp$                                & $D^{*+}\pi^-$                   &  2737.0$\pm$3.5$\pm$11.2      &    73.2$\pm$13.4$\pm$25.0   &                & 2013       & LHCb~\cite{Aaij:2013sza}  \\
 $D^*_{1}(2760)$           & $B^-\rightarrow D^{*0}_{1}\pi^+$     & $D^+\pi^-$                      &  2781$\pm$18$\pm$11$\pm$6     &    177$\pm$32$\pm$20$\pm$7  & $1^-$          & 2015       & LHCb~\cite{Aaij:2015vea}   \\
 $D^*_{3}(2760)$                                      & $B^0\rightarrow D^{*-}_{3}\pi^+$     & $\bar{D}^0\pi^-$                &  2800$\pm$7$\pm$5$\pm$4       &    130$\pm$16$\pm$7$\pm$12  & $3^-$          & 2015       & LHCb~\cite{Aaij:2015sqa}   \\
                           & $e^+e^-$                             & $D^+\pi^-$, $D^{*+}\pi^-$       &  2763.3$\pm$2.3$\pm$2.3       &    60.9$\pm$5.1$\pm$3.6     &                & 2010       & \emph{BABAR}~\cite{delAmoSanchez:2010vq}   \\
                           &  $pp$                                & $D^{*+}\pi^-$                   &  2760.1$\pm$1.1$\pm$3.7       &    74.4$\pm$3.4$\pm$19.1    &                & 2013       & LHCb~\cite{Aaij:2013sza}  \\

 $D^*_{s1}(2700)$          & $B^+\rightarrow \bar{D}^0D^{*+}_{s1}$& $D^0K^+$                        &  2708$\pm$9$^{+11}_{-10}$     &    108$\pm$23$^{+36}_{-31}$ & $1^-$          & 2008       & Belle~\cite{Brodzicka:2007aa}  \\
                           & $B^{+(0)}\rightarrow \bar{D}^0(D^-)D^{*+}_{s1}$& $D^0K^+$  &  2699$^{+14}_{-7}$            &    127$^{+24}_{-19}$        & $1^-$          & 2015       & \emph{BABAR}~\cite{Lees:2014abp}     \\
                           & $e^+e^-$                             & $D^0K^+$, $D^+K^0_s$            &  2688$\pm$4$\pm$4             &    112$\pm$7$\pm$36         &                & 2006       & \emph{BABAR}~\cite{Aubert:2006mh}   \\
                           &  $e^+e^-$                            & $D^{(*)0}K^+$, $D^{(*)+}K^0_s$  &  2710$\pm$2$^{+12}_{-7}$      &    149$\pm$7$^{+39}_{-52}$  &                & 2009       & \emph{BABAR}~\cite{Aubert:2009ah}\\
                           &  $pp$                                & $D^0K^+$, $D^+K^0_s$            &  2709.2$\pm$1.9$\pm$4.5       &    115.8$\pm$7.3$\pm$12.1   &                & 2012       & LHCb~\cite{Aaij:2012pc}  \\
 $D^*_{s1}(2860)$          & $B_s^0\rightarrow D^{*-}_{s1}\pi^+$  & $\bar{D}^0K^-$                  &  2859$\pm$12$\pm$6$\pm$23     &    159$\pm$23$\pm$27$\pm$72 & $1^-$          & 2014       & LHCb~\cite{Aaij:2014baa, Aaij:2014xza}     \\
 $D^*_{s3}(2860)$          & $B_s^0\rightarrow D^{*-}_{s3}\pi^+$  & $\bar{D}^0K^-$                  &  2860.5$\pm$2.6$\pm$2.5$\pm$6.0&    53$\pm$7$\pm$4$\pm$6   & $3^-$           & 2014       & LHCb~\cite{Aaij:2014baa, Aaij:2014xza}     \\
                           & $e^+e^-$                             & $D^0K^+$, $D^+K^0_s$            &  2856.6$\pm$1.5$\pm$5.0       &    48$\pm$7$\pm$10          &                & 2006       & \emph{BABAR}~\cite{Aubert:2006mh}   \\
                           & $e^+e^-$                             & $D^{(*)0}K^+$, $D^{(*)+}K^0_s$  &  2862$\pm$2$^{+5}_{-2}$       &    48$\pm$3$\pm$6           &                & 2009       & \emph{BABAR}~\cite{Aubert:2009ah}   \\
                           &  $pp$                                & $D^0K^+$, $D^+K^0_s$            &  2866.1$\pm$1.0$\pm$6.3       &    69.9$\pm$3.2$\pm$6.6     &                & 2012       & LHCb~\cite{Aaij:2012pc}  \\

\bottomrule[1pt]\bottomrule[1pt]
\end{tabular*}
\label{table1}
\end{table*}

With experimental progress, more and more heavy-light mesons were
reported (see Refs. \cite{Aubert:2003fg, Besson:2003cp, Abe:2003jk,
Krokovny:2003zq, Aubert:2003pe, Aubert:2004pw, Aubert:2006bk,
Aubert:2006nm, Choi:2015uga, Link:2003bd, Abe:2003zm, Aubert:2009wg,
Aubert:2006zb, Aaij:2015sqa, Abazov:2007vq, Aaltonen:2008aa,
Aaij:2015qla, Aaltonen:2007ah, Abazov:2007af, Aaij:2012uva,
Aubert:2006mh, Aubert:2009ah, delAmoSanchez:2010vq, Aaij:2012pc,
Aaij:2013sza, Brodzicka:2007aa, Lees:2014abp, Aaij:2014baa,
Aaij:2015vea, Aaltonen:2013atp, Aaij:2014xza}), which can be grouped
as the candidates of radial and orbital excitations of heavy-light
meson family. In the narrow mass range $2.6-2.9$ GeV, there exists
an accumulation of a large number of observed charmed and
charmed-strange mesons. The information of these states is collected
into Table \ref{table1} for the convenience of readers. Before the
present paper, the strong decay behaviors of these states have been
extensively studied in the context of various models, e.g., the
pseudoscalar emission decay model \cite{Godfrey:2013aaa}, the
$^3P_0$ decay model
\cite{Close:2005se,Close:2006gr,Zhang:2006yj,Li:2010vx,Sun:2010pg,Song:2014mha,Song:2015nia,Song:2015fha,Li:2009qu,Lu:2014zua,Segovia:2013sxa,Segovia:2015dia,Godfrey:2014fga,Yuan:2012ej,Yu:2014dda},
the chiral quark model \cite{Di
Pierro:2001uu,Zhong:2008kd,Zhong:2010vq}, the effective Lagrangian
approach~\cite{Colangelo:2010te,Colangelo:2012xi,Wang:2010ydc,Wang:2013tka,Wang:2014vra},
and the EHQ' decay formula~\cite{Chen:2011rr}. These theoretical
works gave a big push to the development of the research of
heavy-light meson.

\begin{figure}[htbp]
\begin{center}
\includegraphics[width=8.6cm,keepaspectratio]{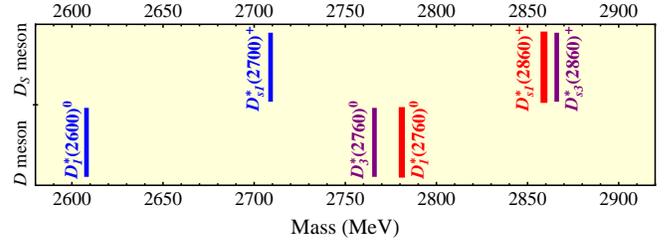}
\caption{$D^*_1(2600)$, $D^*_1(2760)$, and $D_3^*(2760)$ and the
corresponding strange partners $D^*_{s1}(2700)$, $D^*_{s1}(2860)$,
and $D_{s3}^*(2860)$. }\label{Fig1}
\end{center}
\end{figure}

Several experimental measurements of their spin-parity quantum
number $J^P$ were performed recently by the LHCb collaboration,
i.e., two different $D_s$ states, $D^*_{s1}(2860)$ and
$D_{s3}^*(2860)$, were observed in 2014 in the channel of $B_s^0 \to
D_{sJ}^{*-}\pi^+ \to \bar{D}^0K^-\pi^+$, where the obtained helicity
angle distributions indicate that they are the spin-1 and spin-3
states \cite{Aaij:2014baa, Aaij:2014xza}. According to their decay
mode of $\bar{D}^0K^-$, they should be the natural parity mesons,
which satisfy $P = (-1)^J$. In a word, their spin-parity quantum
numbers are $1^-$ and $3^-$, respectively. A spin-1 state,
$D^*_1(2760)^0$, was observed in the channel of $B^- \to
D_1^{*0}\pi^+ \to D^+\pi^-\pi^+$~\cite{Aaij:2015vea}. It is a
possible \emph{D} meson with $J^P=1^-$ . Very recently,
$D^*_3(2760)^-$ was reported in the process of $B^0 \to
D_3^{*-}\pi^+ \to \bar{D}^0\pi^-\pi^+$. Its spin-parity was
determined for the first time as $J^P = 3^-$~\cite{Aaij:2015sqa}.
The new experimental results provide more abundant information for
these states. Until now, four $J^P=1^-$ states and two $J^P=3^-$
open-charmed states, which have natural parity, have been
established\footnote{The spin-parity quantum number of
$D^*_{s1}(2700)$ has been determined as $J^P=1^-$ by
Belle~\cite{Brodzicka:2007aa} and \emph{BABAR}~\cite{Lees:2014abp}.
According to the mass and decay behaviors, the LHCb collaboration
tentatively identified $D^*_1(2600)$ as the $2S~(1^-)$
state~\cite{Aaij:2013sza}.} (see Fig. \ref{Fig1}). They can be
categorized into the $2S$ and $1D$ $D$ and $D_s$ states, where
$D^*_1(2600)$, $D^*_1(2760)$, and $D_3^*(2760)$ are the nonstrange
partners of $D^*_{s1}(2700)$, $D^*_{s1}(2860)$, and
$D_{s3}^*(2860)$, respectively. Their mass gaps reflect the
similarity between charmed and charmed-strange meson families well.

These observed $D^*_1(2600)$, $D^*_1(2760)$, and $D_3^*(2760)$ and
their strange partners $D^*_{s1}(2700)$, $D^*_{s1}(2860)$, and
$D_{s3}^*(2860)$ inspire our interest in carrying out a combined
study of $2S$ and $1D$ open-charm mesons with natural spin-parity,
which is due to the following reasons:
\begin{enumerate}
\item The corresponding studies of these six open-charmed mesons can be borrowed from each other, which is due to the similarity between charmed and charmed-strange meson families.

\item The present study is involved in an intriguing issue of $2S$-$1D$ mixing. Rosner discussed the $2S$-$1D$ mixing effect for two charmonia $\psi(3770)$ and
$\psi(3686)$ \cite{Rosner:2001nm}. For these involved open-charmed
mesons in this paper, there should also exist $2S$-$1D$ mixing when
studying $D^*_1(2600)/D^*_1(2760)$ and
$D^*_{s1}(2700)/D^*_{s1}(2860)$.

\item Besides giving the discussion of $2S$-$1D$ mixing effect existing in these $2S$ and $1D$ open-charm mesons, the main task of this paper is to establish $2S$ and $1D$ open-charm mesons with natural spin-parity, which is also helpful to test whether there exists the spin-orbit inversion in the
heavy-light mesons \cite{Schnitzer:1978kf,Isgur:1998kr} or not
\cite{Woo Lee:2006wm}. Before giving a definite answer, it is
obvious that establishing the $2S$ and $1D$ open-charm mesons with
natural spin-parity is a key step.
\end{enumerate}

Just considering the present update experimental status of $2S$ and
$1D$ open-charm mesons with natural spin-parity and the importance
of studying them just mentioned above, in this paper, we are
dedicated to this interested research topic by considering
$D^*_1(2600)$, $D^*_1(2760)$, and $D_3^*(2760)$ and their strange
partners $D^*_{s1}(2700)$, $D^*_{s1}(2860)$, and $D_{s3}^*(2860)$,
where we mainly focus on their two-body Okubo-Zweig-Iizuka allowed
strong decay behaviors, which can provide valuable information of
their features including partial and total decay widths. For
calculating these decays, the Eichten-Hill-Quigg (EHQ) decay formula
\cite{Eichten:1993ub} is adopted in this paper, which is an
effective approach to deal with the strong decay calculation
involved in the open-charm meson \cite{Eichten:1993ub}.

The manuscript is arranged as follows. In Sec. \ref{sec2}, we first
give a brief introduction to the EHQ' formula. Then, we give the
further phenomenological analysis combined with the present
experimental data. In Sec. \ref{sec3}, the paper ends with the
discussion and conclusion.

\section{The decay behaviors of the discussed $2S$ and $1D$ open-charm mesons with natural spin-parity}\label{sec2}

\begin{table*}[htbp]
\caption{The values of $\mathcal {C}^{s_Q,j'_q,J'}_{j_h,J,j_q}$ for
the different decay modes of these discussed open-charm mesons with
the $2^3S_1$, $1^3D_1$, and $1^3D_3$ quantum numbers. Here,
$\mathcal {P}$ and $\mathcal{V}$ in the first row denote the light
pseudoscalar and vector mesons, respectively. In addition, $s$, $p$,
$d$, $f$, and $g$  in the brackets represent that the corresponding
decays occur via $S$-wave, $P$-wave, $D$-wave, $F$-wave, and
$G$-wave, respectively. }\label{table2}
\renewcommand\arraystretch{1.5}
\begin{center}
\begin{tabular*}{175mm}{c@{\extracolsep{\fill}}ccccccc}
\toprule[1pt]\toprule[1pt]
 $nL(J^P)$                 & $1S(0^-)$~+~$\mathcal {P}$  & $1S(1^-)$~+~$\mathcal {P}$    &  $1S(0^-)$~+~$\mathcal {V}$   &  $1P(1^+,1/2)$+$\mathcal {P}$   &  $1P(1^+,3/2)$+$\mathcal {P}$           &  $1P(2^+)$~+~$\mathcal {P}$            &  $2S(0^-)$~+~$\mathcal {P}$ \\
\midrule[1pt]
 $2^3S_1$                  & $\sqrt{1/3}~(p)$            & $-\sqrt{2/3}~(p)$             & $\sqrt{1/3}~(p)$              & 1~(\emph{s})                    & $\sqrt{1/2}$~(\emph{d})                 & $-\sqrt{3}/2$~(\emph{d})            & $\cdots$  \\
 $1^3D_1$                  & $-\sqrt{2/3}~(p)$           & $-\sqrt{1/3}~(p)$             & $-\sqrt{2/3}~(p)$             & 1~(\emph{d})                    & $-$1~(\emph{s}),~~$-\sqrt{1/2}$~(\emph{d})& $-\sqrt{1/2}$~(\emph{d})               & $-\sqrt{2/3}~(p)$  \\
 $1^3D_3$                  & $\sqrt{3/7}~(f)$            & $-\sqrt{4/7}~(f)$             & $\sqrt{3/7}~(f)$              & $-\sqrt{4/7}~(d)$               & $\sqrt{1/7}~(d)$,~~$\sqrt{9/14}$~(\emph{g})&$-\sqrt{6/7}$~(\emph{d}),~~-$\sqrt{5/14}$~(\emph{g}) & $\sqrt{3/7}~(f)$     \\
\bottomrule[1pt]\bottomrule[1pt]
\end{tabular*}
\end{center}
\end{table*}

\subsection{EHQ' decay formula}
As an approximate symmetry existing in the heavy-light meson system,
the heavy quark symmetry (HQS) plays an important role in the
dynamics of the heavy-light meson. In the HQS limit $(m_Q\to
\infty)$, only the light component of a heavy-light meson takes an
active part in the strong decays \cite{Isgur:1991wq}. Under this
consideration, two heavy-light mesons, with the same light degrees
of freedom $j_q$, but different total angular momentum \emph{J},
shall have the similar decay properties. In general, these two
states can be grouped in one doublet. This picture is supported by
the available properties of \emph{P}-wave heavy-light
mesons~\cite{Agashe:2014kda}.

Based on this picture mentioned above, a decay formula which depicts
the transition between two heavy-light mesons was proposed by
Eichten, Hill, and Quigg in Ref. \cite{Eichten:1993ub}, which is
also called as the EHQ' formula, i.e.,
\begin{equation}
\Gamma^{H\rightarrow H^\prime h}_{j_h,\ell} = \xi\,\left(\mathcal
{C}^{s_Q,j'_q,J'}_{j_h,J,j_q}\right)^2\left|\mathcal
{M}^{j_q,j'_q}_{j_h,\ell}(q/\beta)\right|^2 \,q\,
e^{-q^2/(6\beta^2)}, \label{eq1}
\end{equation}
where $\xi$ is the flavor factor which can be found in
Ref.~\cite{Chen:2012zk}. $q=|\vec q|$ denotes the three-momentum of
the final state in the rest frame of the initial state. \emph{H} and
$H'$ represent the initial and final heavy-light mesons,
respectively. $h$ denotes the light flavor meson. In addition,
$\mathcal {C}^{s_Q,j'_q,J'}_{j_h,J,j_q}$ is a normalized
coefficient, which satisfies the following relation:
\begin{eqnarray}\label{eq2}
\mathcal {C}^{s_Q,j'_q,J'}_{j_h,J,j_q}=\sqrt{(2J'+1)(2j_q+1)}\left\{
           \begin{array}{ccc}
                    s_Q  & j'_q & J'\\
                    j_h  & J    & j_q\\
                    \end{array}
     \right\},\label{h1}
\end{eqnarray}
where $\vec{j}_h \equiv \vec{s}_h + \vec{\ell}$. The symbols $s_h$
and $\ell$ represent the spin of the light meson $h$ and the orbital
angular momentum relative to $H'$, respectively. The transition
factors $\mathcal {M}^{j_q,j'_q}_{j_h,\ell}(p/\beta)$ which involve
the concrete dynamics can be calculated by phenomenological models,
like the relativistic chiral quark model~\cite{Goity:1998jr} and the
$^3P_0$ model~\cite{Page:1998pp,Chen:2011rr}. Equation (\ref{h1})
reflects the requirement of HQS. Until now, the EHQ' formula has
been applied to study open-charm mesons
\cite{Chen:2012zk,Page:1998pp,Chen:2011rr} and excited heavy
baryons~\cite{Chen:2014nyo}.

\begin{table}[htbp]
\caption{The transition factors $\mathcal
{M}^{j_q,j'_q}_{j_h,l}(p/\beta)$ for different decay channels of the
discussed $2^3S_1$, $1^3D_1$, and $1^3D_3$ open-charm mesons. Here,
we define $\mathcal {G} = 32\pi^{1/4}\gamma /(9\beta^{1/2})$.
}\label{table3}
\renewcommand\arraystretch{1.5}
\begin{tabular*}{82mm}{l@{\extracolsep{\fill}}c}
\toprule[1pt]\toprule[1pt]
 $n^{2S+1}L_J\rightarrow n'L'(j'^{P'}_\ell)+ \mathcal
{\mathcal {H}}$  & $\mathcal
{M}^{j_q,j'_q}_{j_h,\ell}(p/\beta)$\\
\midrule[1pt]
$2^3S_1\rightarrow 1S(\frac{1}{2}^-)+ 0^-$   & $-\mathcal {G} \frac{5}{3^2}\frac{q}{\beta}(1-\frac{2}{15}\frac{q^2}{\beta^2})$  \\
$2^3S_1\rightarrow 1S(\frac{1}{2}^-)+ 1^-$   & $-\mathcal {G} \frac{2^{1/2}\cdot5}{3^2}\frac{q}{\beta}(1-\frac{2}{15}\frac{q^2}{\beta^2})$  \\
$2^3S_1\rightarrow 1P(\frac{1}{2}^+)+ 0^-$   & $-\mathcal {G} \frac{1}{2^{1/2}\cdot3^{3/2}}(1-\frac{7}{9}\frac{q^2}{\beta^2}+\frac{2}{27}\frac{q^4}{\beta^4})$   \\
$2^3S_1\rightarrow 1P(\frac{3}{2}^+)+ 0^-$   & $\mathcal {G} \frac{13}{3^{7/2}}\frac{q^2}{\beta^2}(1-\frac{2}{39}\frac{q^2}{\beta^2})$  \\

$1^3D_1\rightarrow 1S(\frac{1}{2}^-)+ 0^-$   & $\mathcal {G} \frac{5^{1/2}\cdot2^{1/2}}{3^2}\frac{q}{\beta}(1-\frac{2}{15}\frac{q^2}{\beta^2})$ \\
$1^3D_1\rightarrow 1S(\frac{1}{2}^-)+ 1^-$   & $-\mathcal {G} \frac{2^{1/2}}{3^2}\frac{q}{\beta}(1-\frac{2}{15}\frac{q^2}{\beta^2})$\\
$1^3D_1\rightarrow 1P(\frac{1}{2}^+)+ 0^-$   & $\mathcal {G} \frac{5^{1/2}}{3^{7/2}}\frac{q^2}{\beta^2}(1+\frac{1}{9}\frac{q^2}{\beta^2})$ \\
$1^3D_1\rightarrow 1P(\frac{3}{2}^+)+ 0^-$   & $-\mathcal {G} \frac{2\cdot5^{1/2}}{3^{3/2}}(1-\frac{5}{18}\frac{q^2}{\beta^2}+\frac{1}{135}\frac{q^4}{\beta^4})$ \\
                                                        & $-\mathcal {G} \frac{13}{3^{7/2}\cdot5^{1/2}}\frac{q^2}{\beta^2}(1-\frac{2}{39}\frac{q^2}{\beta^2})$ \\
$1^3D_1\rightarrow 2S(\frac{1}{2}^-)+ 0^-$   & $\mathcal {G} \frac{5^{1/2}}{3^{7/2}}\frac{q}{\beta}(1-\frac{29}{30}\frac{q^2}{\beta^2}+\frac{1}{45}\frac{q^4}{\beta^4})$ \\

$1^3D_3\rightarrow 1S(\frac{1}{2}^-)+ 0^-$    & $-\mathcal {G} \frac{2^{3/2}}{3^3\cdot5^{1/2}}\frac{1}{\beta^3}$ \\
$1^3D_3\rightarrow 1S(\frac{1}{2}^-)+ 1^-$    & $\mathcal {G} \frac{29^{1/2}}{3^3\cdot5}\frac{1}{\beta^3}$ \\
$1^3D_3\rightarrow 1P(\frac{1}{2}^+)+ 0^-$    & $-\mathcal {G} \frac{2\cdot5^{1/2}}{3^{7/2}}\frac{1}{\beta^2}(1-\frac{1}{15}\frac{q^2}{\beta^2})$ \\
$1^3D_3\rightarrow 1P(\frac{3}{2}^+)+ 0^-$    & $\mathcal {G} \frac{2^{5/2}\cdot7^{1/2}}{3^{7/2}\cdot5^{1/2}}\frac{1}{\beta^2}(1-\frac{1}{42}\frac{q^2}{\beta^2})$ \\
                                              & $\mathcal {G} \frac{2^2}{3^4\cdot5^{1/2}\cdot7^{1/2}}\frac{1}{\beta^4}$ \\
$1^3D_3\rightarrow 2S(\frac{1}{2}^-)+ 0^-$    & $\mathcal {G} \frac{2^2}{3^{5/2}\cdot5^{1/2}}\frac{1}{\beta^3}(1-\frac{1}{36}\frac{q^2}{\beta^2})$ \\
\bottomrule[1pt]\bottomrule[1pt]
\end{tabular*}
\end{table}

\begin{table*}[htbp]
\caption{Decay widths of $D_1^*(2600)$ and $D^*_1(2760)$ with
several typical $\phi$ mixing angles (in MeV).} \label{table4}
\renewcommand\arraystretch{1.3}
\begin{tabular*}{170mm}{@{\extracolsep{\fill}}ccccccccccc}
\toprule[1pt]\toprule[1pt]
Decay  & \multicolumn{5}{c}{$D_1^*(2600)$}  & \multicolumn{5}{c}{$D_1^*(2760)$}   \\
\cline{2-6}\cline{7-11}
channels  & $-30^\circ$  &   $-15^\circ$  &   $0^\circ$   &  $15^\circ$  &  $30^\circ$    &   $-30^\circ$   & $-15^\circ$ &   $0^\circ$ &   $15^\circ$  & $30^\circ$   \\
\cline{1-6}\cline{7-11}
$D~\pi$           &  107.2    &  89.2      &  62.2      &  33.6    &  10.9      &  3.7     &  18.0     &  39.4     &  62.1    &  80.0     \\
$D^*~\pi$         &  48.3     &  84.5      &  116.9     &  136.7   &  138.7     &  96.9    &  58.7     &  24.6     &  3.7     &  1.6     \\
$D_s~K$           &  11.9     &  9.9       &  6.9       &  3.7     &  1.2       &  0.9     &  4.4      &  9.6      &  15.1    &  19.5    \\
$D^*_s~K$         &  0.0      &  0.4       &  1.0       &  1.7     &  2.4       &  11.7    &  7.1      &  3.0      &  0.5     &  0.2  \\
$D~\eta$          &  16.2     &  13.5      &  9.4       &  5.1     &  1.7       &  0.9     &  4.6      & 10.0      &  15.8    &  20.3  \\
$D^*~\eta$        &  2.2      &  3.9       &  5.3       &  6.2     &  6.2       &  15.7    &  9.5      &  4.0      &  0.6     &  0.2   \\
$D~\rho$          &  $\cdots$      &  $\cdots$       &  $\cdots$       &  $\cdots$     &  $\cdots$       &  42.1    &  21.5     &  6.1      &  0.0     &  4.9   \\
$D~\omega$        &  $\cdots$      &  $\cdots$       &  $\cdots$       &  $\cdots$     &  $\cdots$       &  13.2    &  6.7      &  1.9      &  0.0     &  1.6  \\
$D'_1(2430)~\pi$  &  2.1      &  2.6       &  2.8       &  2.6     &  2.1       &  1.6     &  1.0      &  0.4      &  0.1     &  0.0   \\
$D_1(2420)~\pi$   &  37.1     &  11.0      &  0.1       &  7.3     &  30.8      &  121.4   &  165.5    &  191.4    &  192.1   &  167.5  \\
$D_2^*(2460)~\pi$ &  0.0      &  0.0       &  0.0       &  0.0     &  0.0       &  1.1     &  0.4      &  0.0      &  0.1     &  0.5   \\
$D(2550)~\pi$     &  $\cdots$      &  $\cdots$       &  $\cdots$       &  $\cdots$     &  $\cdots$       &  0.0     &  0.1      &  0.1      &  0.2     &  0.3   \\
Total             &  225.0    &  215.0     &  204.6     &  196.9   &  194.0     &  309.2   &  297.5    &  290.4    &  290.3   &  296.6   \\
\cline{2-6}\cline{7-11}
Experiment        &  \multicolumn{5}{c}{$104.5\sim175.9$~\cite{Aaij:2013sza}}     &  \multicolumn{5}{c}{$118\sim236$~\cite{Aaij:2015vea}}   \\
\bottomrule[1pt]\bottomrule[1pt]
\end{tabular*}
\end{table*}


\begin{table*}[htbp]
\caption{Decay widths of $D_{s1}^*(2700)$ and $D_{s1}^*(2860)$ with several typical $\phi$ mixing angles (in
MeV).} \label{table5}
\renewcommand\arraystretch{1.3}
\begin{tabular*}{170mm}{@{\extracolsep{\fill}}ccccccccccc}
\toprule[1pt]\toprule[1pt]
Decay  & \multicolumn{5}{c}{$D_{s1}^*(2700)$}  & \multicolumn{5}{c}{$D_{s1}^*(2860)$}   \\
\cline{2-6}\cline{7-11}
channels  & $-30^\circ$  &   $-15^\circ$  &   $0^\circ$   &  $15^\circ$  &  $30^\circ$    &   $-30^\circ$   & $-15^\circ$ &   $0^\circ$ &   $15^\circ$  & $30^\circ$   \\
\cline{1-6}\cline{7-11}
$D~K$          &  140.9    &  117.2      &  81.7      &  44.1    &  14.3                  &  5.7      &  27.6       &  60.4      &  95.2    &  122.7     \\
$D^*~K$        &  46.7     &  81.8       &  113.1     &  132.3   &  134.3                 &  130.3    &  79.0       &  33.1      &  5.0     &  2.1     \\
$D_s~\eta$     &  20.2     &  16.6       &  11.3      &  5.7     &  1.3                   &  1.3      &  6.1        &  13.3      &  21.0    &  27.0    \\
$D_s^*~\eta$   &  1.6      &  2.7        &  3.7       &  4.3     &  4.3                   &  18.9     &  11.5       &  4.8       &  0.7     &  0.3  \\
$D~K^*$        &  $\cdots$      &  $\cdots$        &  $\cdots$       &  $\cdots$     &  $\cdots$                   &  42.3     &  21.6       &  6.1       &  0.0     &  5.0  \\
Total          &  209.4    &  218.3      &  209.8     &  186.4   &  154.2                 &  198.5    &  145.8      &  117.7     &  121.9   &  157.1   \\
\cline{2-6}\cline{7-11}
Experiment        &  \multicolumn{5}{c}{$103\sim195$~\cite{Aubert:2009ah}}     &  \multicolumn{5}{c}{159$\pm$23$\pm$27$\pm$72~\cite{Aaij:2014baa, Aaij:2014xza}}   \\
\bottomrule[1pt]\bottomrule[1pt]
\end{tabular*}
\end{table*}

For the $2S$ and $1D$ open-charm mesons with natural spin-parity,
the relevant values of $\mathcal {C}^{s_Q,j'_q,J'}_{j_h,J,j_q}$ and
the expressions of the corresponding transition factors are listed
in Tables \ref{table2} and \ref{table3}, respectively. These
transition factors are determined by the $^3P_0$ model. Thus, two
parameters $\beta$ and $\gamma$ are introduced in our calculation.
Here, $\beta$ is the scale of harmonic oscillator wave function
depicting the mesons in the discussed
transitions.\footnote{According to the results in Ref.
\cite{Blundell:1996as}, we adopt a universal $\beta$ value for the
mesons involved in the discussed transitions.} Following
Ref.~\cite{Chen:2011rr}, we take $\beta=0.38$ GeV. The parameter
$\gamma$ denotes the strength of quark pair creation from vacuum. In
this paper, we assume that the charmed and charmed-strange mesons
are governed by a same $\gamma$. As shown in our previous
paper~\cite{Chen:2012zk}, all $P$-wave heavy-light mesons were
explained well in the framework of the EHQ' formula, where $\gamma$
is directly fixed by the width of $D_2^*(2460)^0$, i.e.,
$D_2^*(2460)^0$ is a $1^3P_2$ \emph{D} state~\cite{Agashe:2014kda}.
Its mass and width are $2460.47\pm0.14$ MeV and $47.7\pm0.7$ MeV,
respectively \cite{Amhis:2014hma}. Via Eq. (\ref{eq3}), we determine
$\gamma=0.125$. In the following, we will apply the EHQ' formula
with these determined parameters to calculate the decay behaviors of
$2S$ and $1D$ open-charm mesons with natural spin-parity, which is
the main task of present paper. Finally, it needs to be emphasized
that we adopt experimental average values for the masses of these
discussed $2S$ and $1D$ open-charm mesons with natural spin-parity.
For other open-charm mesons and light flavor mesons involved in
calculation, the corresponding masses are taken from Particle Data
Group (PDG) \cite{Agashe:2014kda}.

\subsection{$D^*_1(2600)$, $D_{s1}^*(2700)$, $D^*_1(2760)$ and $D_{s1}^*(2860)$ as pure states}

First, we discuss $D^*_1(2600)$, $D_{s1}^*(2700)$, $D^*_1(2760)$ and
$D_{s1}^*(2860)$ without introducing $2S$-$1D$ mixing, where their
partial and total decay widths are collected in Tables \ref{table4}
and \ref{table5} (see the results corresponding to mixing angle
$\phi=0^\circ$).

With the $2^3S_1$ assignment, the total decay width of $D_1^*(2600)$
is predicted as 204.6 MeV, which is larger than the averaged value
of widths measured by \emph{BABAR}~\cite{delAmoSanchez:2010vq} and
LHCb~\cite{Aaij:2013sza}, but is close to the upper limit of the
measurement by LHCb~\cite{Aaij:2013sza} (see Table \ref{table4}). We
also predict the branching ratio
\begin{eqnarray}
\frac{\mathcal {B}(D_1^*(2600)\rightarrow D~\pi)}{\mathcal
{B}(D_1^*(2600)\rightarrow D^*~\pi)}~=~0.53, \label{eq3}
\end{eqnarray}
which is close to the upper limit of the \emph{BABAR} result
$(0.21\sim0.43)$ ~\cite{delAmoSanchez:2010vq}. It is obvious that
the main feature of $D_1^*(2600)$ can be understood under the
$2^3S_1$ assignment.

If $D_{s1}^*(2700)$ is the strange partner of $D_1^*(2600)$, the
predicted width can reach up to 209.8 MeV which is near to the upper
limit of the \emph{BABAR} result~\cite{Aubert:2009ah} (see Table
\ref{table5}). Furthermore, the predicted branching ratio is
\begin{eqnarray}
\frac{\mathcal {B}(D^*_{s1}(2700)\rightarrow D^*~K)}{\mathcal
{B}(D^*_{s1}(2700)\rightarrow D~K)}~=~1.43, \label{eq4}
\end{eqnarray}
which is close to the upper limit of the \emph{BABAR} result
$(0.66-1.16)$~\cite{Aubert:2009ah}. The above facts indicate that
$D_{s1}^*(2700)$ has a dominant $2^3S_1$ component.

\begin{figure}[htpb]
\begin{center}
\includegraphics[width=8.6cm,keepaspectratio]{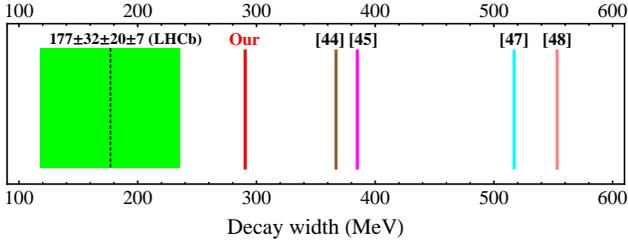}
\caption{A comparison of the experimental \cite{Aaij:2015vea} and
theoretical
widths~\cite{Lu:2014zua,Close:2005se,Zhong:2010vq,Song:2015fha} for
the $D^*_1(2760)$ with the $D(1^3D_1)$ assignment.}\label{Fig2}
\end{center}
\end{figure}

As a $1^3D_1$ state, the calculated total decay width of
$D^*_1(2760)$ is around 300 MeV. Before this paper, the studies in
Refs.~\cite{Lu:2014zua,Close:2005se,Zhong:2010vq,Song:2015fha} also
suggested that the decay width of the $1^3D_1$ \emph{D} state should
be large. We list the theoretical and experimental results in
Fig.~\ref{Fig2} for the comparison. Among these allowed decay modes,
the channels of $D\pi$, $D^*\pi$ and $D_1(2420)\pi$ are important
for $D^*_1(2760)$ due to large partial widths, e.g., the channel of
$D_1(2420)\pi$ with branching ratio $65.9\%$ is its dominant decay
mode. The same conclusion has been obtained in
Refs.~\cite{Lu:2014zua,Close:2005se,Zhong:2010vq,Song:2015fha}. As a
$1^3D_1$ state, the below ratio,
\begin{eqnarray}
\begin{aligned}
\frac{\mathcal {B}(D^*_1(2760)\rightarrow D~\pi)}{\mathcal
{B}(D_1(2760)\rightarrow D^*~\pi)}~=~1.60,\label{eq5}
\end{aligned}
\end{eqnarray}
is predicted for $D^*_1(2760)$, which is consistent with the results
in Refs.~\cite{Lu:2014zua,Close:2005se,Zhong:2010vq,Song:2015fha}
(see Table~\ref{table6}).

\begin{table}[htbp]
\caption{Comparison of different theoretical results for the ratio
of $\mathcal {B}(D^*_1(2760)\rightarrow D~\pi)/\mathcal
{B}(D_1(2760)\rightarrow D^*~\pi)$. }\label{table6}
\renewcommand\arraystretch{1.2}
\begin{tabular*}{85mm}{c@{\extracolsep{\fill}}cccc}
\toprule[1pt]\toprule[1pt]
Ref.~\cite{Lu:2014zua} & Ref.~\cite{Song:2015fha}     & Ref.~\cite{Close:2005se}  & Ref.~\cite{Zhong:2010vq}         & Our\\
\hline
2.04                   & 2.17                         & 1.62                      & 2.42                             & 1.60     \\
\bottomrule[1pt]\bottomrule[1pt]
\end{tabular*}
\end{table}

As the strange partner of $D^*_1(2760)$, the total decay width of
$D_{s1}^*(2860)$ is around $117.7$ MeV, which is consistent with the
measurement given by LHCb~\cite{Aaij:2014baa, Aaij:2014xza} and
comparable with former results in
Refs.~\cite{Godfrey:2013aaa,Zhang:2006yj,Godfrey:2014fga,Segovia:2015dia}
(see Table~\ref{table7}). Our results also show that the $DK$ mode
is one of the main decay channels, which can explain why
$D_{s1}^*(2860)$ was first found in its $DK$ channel
\cite{Aaij:2014baa, Aaij:2014xza}. Additionally, we obtain the ratio
\begin{eqnarray}
\begin{aligned}
\frac{\mathcal {B}(D_{s1}^*(2860)\rightarrow D~K)}{\mathcal
{B}(D_{s1}^*(2860)\rightarrow D^*~K)}~=~1.82\label{eq6},
\end{aligned}
\end{eqnarray}
which is close to 1.92 given by Ref.~\cite{Godfrey:2014fga} and 1.42
by Ref.~\cite{Segovia:2015dia}.

\begin{table}[htbp]
\caption{Comparison of different theoretical results for the decay
width of $D_{s1}^*(2860)$ (in MeV).}\label{table7}
\renewcommand\arraystretch{1.2}
\begin{tabular*}{85mm}{c@{\extracolsep{\fill}}cccc}
\toprule[1pt]\toprule[1pt]
Ref.~\cite{Godfrey:2013aaa} & Ref.~\cite{Zhang:2006yj}     &  Ref.~\cite{Godfrey:2014fga} & Ref.~\cite{Segovia:2015dia}         & Our\\
\hline
   145                       & 132                          &      186.8                  & 153.2                               & 117     \\
\bottomrule[1pt]\bottomrule[1pt]
\end{tabular*}
\end{table}

In this subsection, we mainly focus on $D_1^*(2600)$,
$D_{s1}^*(2700)$, $D^*_1(2760)$ and $D_{s1}^*(2860)$ as pure states
and give the phenomenological analysis, where the information of
their other decay channels can be found in Tables
\ref{table4}$-$\ref{table5}. In fact, there should exist a $2S$-$1D$
mixing effect to these discussed $D^*(2600)$, $D_{s1}^*(2700)$,
$D^*_1(2760)$ and $D_{s1}^*(2860)$, which will be illustrated in the
next subsection.

\subsection{$D_1^*(2600)$ and $D_{s1}^*(2700)$ as orthogonal partners of $D^*_1(2760)$ and $D_{s1}^*(2860)$ respectively}

In reality, $D_1^*(2600)$ should be the orthogonal partner of
$D^*_1(2760)$. There also exists the similar relation for their
strange partners $D_{s1}^*(2700)$ and $D_{s1}^*(2860)$. Thus, we
need to investigate the $2S$-$1D$ mixing effect, where the mixing
scheme satisfies \cite{Li:2009qu}
\begin{eqnarray}
\begin{aligned}
 \left(
           \begin{array}{c}
                     |(SD)_1\rangle_L\\
                     |(SD)_1\rangle_H\\
                    \end{array}
     \right)&=\left(
           \begin{array}{cc}
                    \cos\phi  & -\sin\phi \\
                    \sin\phi  & \cos\phi\\
                    \end{array}
     \right)  \left(
           \begin{array}{c}
                     |2^3S_1\rangle \\
                     |1^3D_1\rangle \\
                    \end{array}
     \right)
\end{aligned}
\end{eqnarray}
with $\phi$ as the mixing angle. The states marked by subscripts $L$
and $H$ are identified as the low-mass and high-mass mixed states,
respectively. For discussing the $2S$-$1D$ mixing effect, we take
several typical $\phi$ values, i.e., $\phi = -30^\circ$,
$-15^\circ$, $15^\circ$, and $30^\circ$. The corresponding partial
and total widths of $D_1^*(2600)$, $D_{s1}^*(2700)$, $D^*_1(2760)$
and $D_{s1}^*(2860)$ are shown in
Tables~\ref{table4}$-$\ref{table5}. Our results indicate that there
exists dependence of the total widths of $D_1^*(2600)$,
$D_{s1}^*(2700)$, and $D_1^*(2760)$ on the mixing angle $\phi$. The
width of $D_{s1}^*(2860)$ highly depends on $\phi$.

We notice that the ratios of partial decay widths are good physical quantities to constrain the mixing angle $\phi$.
At present, \emph{BABAR}~\cite{delAmoSanchez:2010vq,Aubert:2009ah} measured ratios
\begin{eqnarray}
\frac{\mathcal {B}(D^*_1(2600)\rightarrow D~\pi)}{\mathcal
{B}(D^*_1(2600)\rightarrow D^*~\pi)}=0.32\pm0.02\pm0.09, \label{eq1}
\end{eqnarray}
and
\begin{eqnarray}
\frac{\mathcal {B}(D^*_{s1}(2700)\rightarrow D^*~K)}{\mathcal
{B}(D^*_{s1}(2700)\rightarrow D~K)}=0.91\pm0.13\pm0.12. \label{eq2}
\end{eqnarray}
If fitting these two ratios, the mixing angles
$\phi=4^\circ-17^\circ$ and $\phi=-4^\circ$ to $-16^\circ$ are
obtained for $D^*_1(2600)$ and $D_{s1}^*(2700)$, respectively, which
show that our results by the EHQ' formula favor small mixing angle.

Besides the $D^{(*)}\pi$ and $D^{(*)}K$ modes, the $D_{(s)}\eta$ and
$D_sK$ channels are also important to reflect the features of these
discussed open-charm mesons. Furthermore, we get the following
ratios:
\begin{eqnarray*}
\begin{aligned}
&\frac{\mathcal {B}(D^*_1(2600)\rightarrow
D~\eta)}{\mathcal {B}(D^*_1(2600)\rightarrow D~\pi)}~=~0.15,\\
&\frac{\mathcal {B}(D^*_1(2600)\rightarrow
D_s~K)}{\mathcal {B}(D^*_1(2600)\rightarrow D~\pi)}~=~0.11,\\
&\frac{\mathcal {B}(D_{s1}^*(2700)\rightarrow D_s~\eta)}{\mathcal
{B}(D_{s1}^*(2700)\rightarrow D~K)}~=~0.14,\\
&\frac{\mathcal {B}(D_{s1}^*(2860)\rightarrow D_s~\eta)}{\mathcal
{B}(D_{s1}^*(2860)\rightarrow D~K)}~=~0.22.
\end{aligned}
\end{eqnarray*}
We need to specify that these four ratios are weakly dependent on
the mixing angle, and independent on the parameter $\gamma$.

If the $2S$-$1D$ mixing effect exists, $D^*_1(2760)/D_{s1}^*(2860)$
should be the orthogonal partner of $D^*_1(2600)/D_{s1}^*(2700)$.
With the ranges of mixing angle obtained above, we predict
\begin{eqnarray}
\frac{\mathcal {B}(D^*_1(2760)\rightarrow D~\pi)}{\mathcal
{B}(D^*_1(2760)\rightarrow D^*~\pi)} = 2.62-28.86, \label{eq10}
\end{eqnarray}
and
\begin{eqnarray}
\frac{\mathcal {B}(D^*_{s1}(2860)\rightarrow D^*~K)}{\mathcal
{B}(D^*_{s1}(2860)\rightarrow D~K)} = 0.31-1.16. \label{eq11}
\end{eqnarray}
which are different from the results in Eqs.~(\ref{eq5}) and
(\ref{eq6}), respectively. These results can be also applied to test
the $2S$-$1D$ mixing effect.


\begin{table}[htbp]
\caption{The partial and total widths of $D^*_3(2760)$ and
$D_{s3}^*(2860)$ as $D(1^3D_3)$ and $D_s(1^3D_3)$, respectively.
}\label{table8}
\renewcommand\arraystretch{1.1}
\begin{tabular*}{85mm}{l@{\extracolsep{\fill}}ccccc}
\toprule[1pt]\toprule[1pt]
      \multicolumn{4}{c}{$D^*_3(2760)$}  &              \multicolumn{2}{c}{$D_{s3}^*(2860)$} \\
\cline{1-4}\cline{5-6}
 Modes        & $\Gamma_i$ (MeV)  & Modes             & $\Gamma_i$ (MeV) & Modes          & $\Gamma_i$ (MeV)\\
\cline{1-4}\cline{5-6}
$D~\pi$       & 27.9             & $D~\omega$         & 0.1             & $D~K$            & 28.5\\
$D_s~K$       & 1.6              & $D~\rho$           & 0.2             & $D^*~K$          & 12.2\\
$D^*~\pi$     & 15.5             & $D'_1(2430)\pi$    & 1.1             & $D_s~\eta$       & 1.9\\
$D_s^*~K$     & 0.2              & $D_1(2420)\pi$     & 0.4             & $D^*_s~\eta$     & 0.4\\
$D~\eta$      & 1.4              & $D^*_2(2460)\pi$   & 1.1             & $D~K^*$          & 0.2\\
$D^*~\eta$    & 0.2              & $D(2550)\pi$       & 0.0             &                  &    \\
\cline{3-4}\cline{5-6}
             &                  & Total        & 49.7            & Total           & 43.2\\
\bottomrule[1pt]\bottomrule[1pt]
\end{tabular*}
\end{table}

\subsection{$D^*_3(2760)$ and $D_{s3}^*(2860)$}

In the following, the phenomenological analysis of two $3^-$
open-charm mesons $D^*_3(2760)$ and $D_{s3}^*(2860)$ is given. The
corresponding partial and total widths are shown in Table
\ref{table8}.

Under the $D(1^3D_3)$ assignment, the predicted widths of
$D^*_3(2760)$ is about 47.9 MeV, which is comparable with the
experimental data reported by
\emph{BABAR}~\cite{delAmoSanchez:2010vq} and
LHCb~\cite{Aaij:2013sza}. The results of Refs.
\cite{Zhong:2010vq,Li:2010vx,Lu:2014zua,Song:2015fha,Chen:2011rr}
also supported $D^*_3(2760)$ as the $D(1^3D_3)$ state.

\begin{figure}[htpb]
\begin{center}
\includegraphics[width=8.6cm,keepaspectratio]{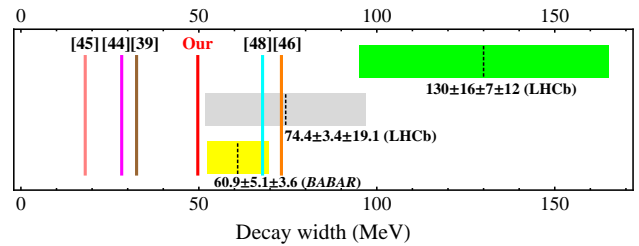}
\caption{A comparison of the
experimental~\cite{delAmoSanchez:2010vq,Aaij:2013sza,Aaij:2015sqa}
and theoretical
widths~\cite{Song:2015fha,Lu:2014zua,Sun:2010pg,Zhong:2010vq,Li:2010vx}
for the $D_3^*(2760)$ with the $D(1^3D_3)$ assignment.}\label{Fig3}
\end{center}
\end{figure}

We notice a new result of $D^*_3(2760)$ released by LHCb very recently \cite{Aaij:2015sqa}, where
the averaged width from two models (the Isobar and K-matrix
formalisms) is \cite{Aaij:2015sqa}
\begin{eqnarray*}
\Gamma(D^*_3(2760))=130\pm16\pm7\pm12~\rm {MeV},
\end{eqnarray*}
which is far larger than the present theoretical
results~\cite{Song:2015fha,Zhong:2010vq,Li:2010vx,Lu:2014zua,Chen:2011rr}
and former experimental data
\cite{delAmoSanchez:2010vq,Aaij:2013sza}. Additionally, the mass of
$D^*_3(2760)$ in Ref. \cite{Aaij:2015sqa} is about 40 MeV higher
than previous experimental data (see Table \ref{table1} for more
details). We list the experimental and theoretical widths in Fig.
\ref{Fig3} for a comparison. The theoretical results favor the
experimental measurements from \emph{BABAR} and LHCb
\cite{delAmoSanchez:2010vq,Aaij:2013sza}. The LHCb collaboration may
overestimate the width of $D^*_3(2760)$ in the latest
measurement~\cite{Aaij:2015sqa}. In the near future, more precise
measurement of the resonant parameter of $D_3^*(2760)$ is crucial to
clarify this mess. The possible explanation for the change of the
new result of $D_3^*(2760)$ given by LHCb \cite{Aaij:2015sqa} will
be presented in the next section.

\begin{figure}[htbp]
\begin{center}
\includegraphics[width=8.6cm,keepaspectratio]{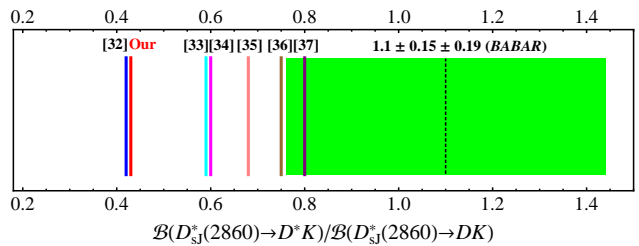}
\caption{A comparison of the experimental~\cite{Aubert:2009ah} and
theoretical
results~\cite{Godfrey:2013aaa,Zhang:2006yj,Li:2009qu,Godfrey:2014fga,Segovia:2015dia,Song:2015nia}
for the ratio $\mathcal {B}(D_{sJ}^*(2860)\rightarrow D^*K)/\mathcal
{B}(D_{sJ}^*(2860)\rightarrow DK)$.}\label{Fig4}
\end{center}
\end{figure}

With the $D_s(1^3D_3)$ assignment, we show the decay behavior of
$D_{s3}^*(2860)$ in Table \ref{table8}, where the calculated total
decay width is about 43.2 MeV, which is in good agreement with the
experimental
data~\cite{Aubert:2006mh,Aubert:2009ah,Aaij:2012pc,Aaij:2014baa,
Aaij:2014xza} (see Table~\ref{table1}) and comparable with other
theoretical results obtained by different phenomenological models
~\cite{Godfrey:2013aaa,Zhang:2006yj,Li:2009qu,Godfrey:2014fga,Segovia:2015dia,Song:2015nia},
where the width of the $D_s(1^3D_3)$ state was predicted in the
range of $14-85$ MeV. However, we also notice that most predicted
values for ratio $\mathcal {B}(D_s^*(1^3D_3)\rightarrow
D^*K)/\mathcal {B}(D_s^*(1^3D_3)\rightarrow DK)$ are smaller than
the \emph{BABAR} measurement~\cite{Aubert:2009ah} (see Fig.
\ref{Fig4}). This discrepancy between theoretical and experimental
results should be clarified by further experimental and theoretical
efforts. In the next section, we will give some discussions for this
point.

\section{Discussion and conclusion}\label{sec3}

In this paper, we carried out a combined study of $2S$ and $1D$
open-charm mesons with natural spin-parity, which was mainly
inspired by the recent new experimental progresses on open-charm
mesons reported by LHCb \cite{Aaij:2014baa, Aaij:2014xza,
Aaij:2015vea, Aaij:2015sqa}. Our study is helpful to identify the
properties of these observed $2S$ and $1D$ open-charm mesons with
natural spin-parity. Besides comparing our results with the present
data, we also gave some typical ratios of partial decay widths and
partial decay widths, which are crucial information to establish
$2S$ and $1D$ open-charm mesons with natural spin-parity. In
addition, the $2S$-$1D$ mixing effect existing in these $2S$ and
$1D$ open-charm mesons with natural spin-parity was discussed, which
is an interested research topic for the experiments in the future.

Before closing this paper, we still want to give more discussions
relevant to these $2S$ and $1D$ open-charm mesons with natural
spin-parity. Checking these collected experimental data in Table
\ref{table1}, we notice that the measurements of the resonant
parameter for the same state by different experiments are different
from each other. Thus, the present crucial task for experiment is to
precisely measure the resonant parameters of these $2S$ and $1D$
open-charm mesons with natural spin-parity. With more experimental
data collected by LHCb and forthcoming Belle II, we believe that
experiments will make more progress on this point.

\begin{figure}[htpb]
\begin{center}
\includegraphics[width=8.6cm,keepaspectratio]{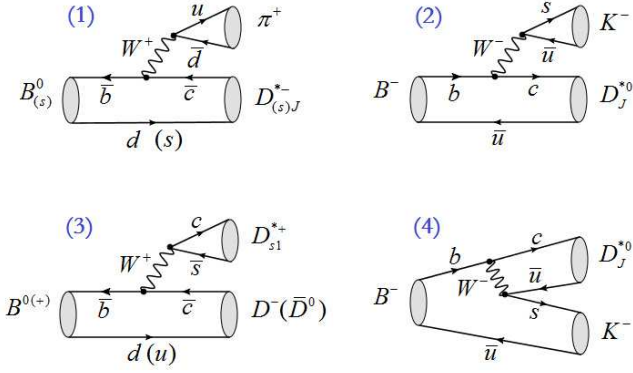}
\caption{The topological diagrams for the productions of excited
\emph{D} and $D_s$ mesons via the nonleptonic decays of $B_{(s)}$
mesons.}\label{Fig5}
\end{center}
\end{figure}


In Fig. \ref{Fig5}, we present the possible topological diagrams
relevant to the productions of these discussed open-charm mesons
through the nonleptonic decays of $B_{(s)}$ mesons. Two states,
$D_{s1}^*(2860)^-$ and $D_{s3}^*(2860)^-$, were observed in the
process depicted in Fig. \ref{Fig5} (1). With the same topological
diagram, however, only $D_3^*(2760)^-$ was reported by LHCb. In
fact, there does not exist any suppression to the $D_1^*(2760)$
state produced via Fig. \ref{Fig5} (1), where the total angle
momentum of $D_1^{*-}(2760)$ is smaller than that of
$D_3^*(2760)^-$. In Ref. \cite{Aaij:2015sqa}, LHCb admitted that
``The current analysis does not preclude a charged spin-1 $D^*$
state at around the same mass.'' Thus, there exists the possibility
that the $D_3^*(2760)^-$ structure reported by LHCb might be from a
superposition of $1^-$ and a $3^-$ $1D$-wave charmed mesons, which
provides a possible explanation of why the experimental width of
$D_3^*(2760)^-$ by LHCb~\cite{Aaij:2015sqa} is obviously larger than
former measurements by the $e^+e^-$ and $pp$ collisions
\cite{delAmoSanchez:2010vq,Aaij:2013sza}.


By $B^-\to D_J^{^*0}(D^+\pi^-)K^-$ corresponding to Fig.~\ref{Fig5}
(2) and (4), where the later one is a color-suppressed process, LHCb
has reported the signal of $D_1^*(2760)^0$, where $D_3^*(2760)^0$ is
missing. According to the experience of the observations of
$D^*_{s1}(2860)$ and $D_{s3}^*(2860)$ in $B_s^0\to
\bar{D}^0K^-\pi^+$ via Fig. \ref{Fig5} (1), there probably exists a
$D_3^*(2760)^0$ signal associated with $D_1^*(2760)^0$ in $B^-\to
D^+\pi^-K^-$ since Fig. \ref{Fig5} (2) has the same topological
structure as that of Fig. \ref{Fig5} (1).


As shown in Tables \ref{table4} and \ref{table6}, the $D_1(2420)\pi$
mode is an important decay channel for $D_{1}^*(2760)$, but a
subordinate channel for $D_3^*(2760)$. Thus, we suggest the
experiment to carry out the analysis of $B^0\to
D_{1}^*(2760)^-\pi^+\to D_{1}(2420)^0\pi^-\pi^+$ and $B^-\to
D_{1}^*(2760)^0K^-\to D_{1}(2420)^+\pi^-K^-$, by which the
$D^*_1(2760)$ signal can be easily disentangled from the
$D^*_3(2760)$ signal  since  the $D^*_3(2760)$ signal is suppressed
here.

As illustrated in Sec. \ref{sec2}, the experimental data of ratio
$\mathcal {B}(D^*~K)/\mathcal {B}(D~K)$ of $D_{s3}^*(2860)$ is
larger than most theoretical
results~\cite{Godfrey:2013aaa,Zhang:2006yj,Li:2009qu,Godfrey:2014fga,Segovia:2015dia,Song:2015nia,Colangelo:2010te}.
The nonstrange partner of $D_{s3}^*(2860)$ corresponds to
$D_3^*(2760)$. We notice that there exists a possible $2^-$ state
$D_2(2750)$~\cite{Aaij:2013sza} (see Table \ref{table1} for more
details). Due to the similarity between charmed and charmed-strange
meson families, we conjugate the existence of a $2^-$ state with
similar mass to that of $D_{s3}^*(2860)$. At present, it is
difficult to exclude the possibility that the present
$D_{s3}^*(2860)$ signal in the $D^*K$ invariant mass spectrum
\cite{Aubert:2009ah} contains a $D_s(2^-)$ structure. If it is true,
the discrepancy between theoretical and experimental results can be
understood well for the ratio $\mathcal {B}(D^*~K)/\mathcal
{B}(D~K)$ of $D_{s3}^*(2860)$~\cite{Chen:2011rr,Colangelo:2012xi},
i.e.,  $\mathcal {B}(D^*~K)/\mathcal {B}(D~K)$ of $D_{s3}^*(2860)$
is overestimated probably.


It is not the end of the story of the study of $2S$ and $1D$
open-charm mesons with natural spin-parity, since there still exist
some puzzles just discussed above, which are waiting for the
solutions given by the joint effort from experimentalists and
theorists.

\begin{acknowledgments}
B. C. would like to thank the Academic Exchange Platform of
Theoretical Physics (Lanzhou University) for supporting his stay at
Lanzhou University. This project is supported by the National
Natural Science Foundation of China under Grants No. 11305003, No.
No. 11222547, No. 11175073, No. 11447604 and No. 11475111, the Key
Program of the He'nan Educational Committee of China under Grant No.
13A140014, the Ministry of Education of China (SRFDP under Grant No.
2012021111000), and the Fok Ying Tung Education Foundation (Grant
No. 131006), and the Innovation Program of Shanghai Municipal
Education Commission under Grant No. 13ZZ066.
\end{acknowledgments}

\end{document}